\documentstyle[epsf,12pt]{article}
\begin{document}
\title{Protons, Deuterons and Flow}
\date{BNL-CERN-Columbia-Hiroshima-LANL-LUND-Nantes-\\
NBI-OSU-Texas A\&M-SUNY-Zagreb}
\pagestyle{empty}
\author{Michael Murray for the NA44 Collaboration}
\maketitle
\begin{abstract}
Proton and deuteron $m_T$ spectra and rapidity densities
at $y=1.9-2.3$ are presented 
for central $SS$, $SPb$ and $PbPb$ collisions. The inverse slopes 
increase with system size and mass.
The  radius parameter, $R_G$, extracted from ratios of the spectra 
increases with system size. 
$R_G$ $\propto 1/\sqrt{m_T}$ consistent with transverse flow and 
NA44's HBT results.
\end{abstract}

\section*{Experiment NA44}
NA44 measures particle spectra and correlations, \cite{mtna44}.
It triggers on rare particles
and has excellent momentum resolution of \( \sigma_{p}/p = 0.2\% \).
The  mean occupancy is 
about 0.5 particles per central $PbPb$
collision. Changing angles
and momentum settings scans 
transverse mass,  $m_T \equiv \sqrt{p_T^2 +m^2}$ and
rapidity.
This paper 
compares deuterons to protons at  the same velocity.

\section*{Results}
\begin{figure}[h]
  \vspace{-0.5cm}
 \begin{center}
    \mbox{
     \epsfxsize=11cm
     \epsffile{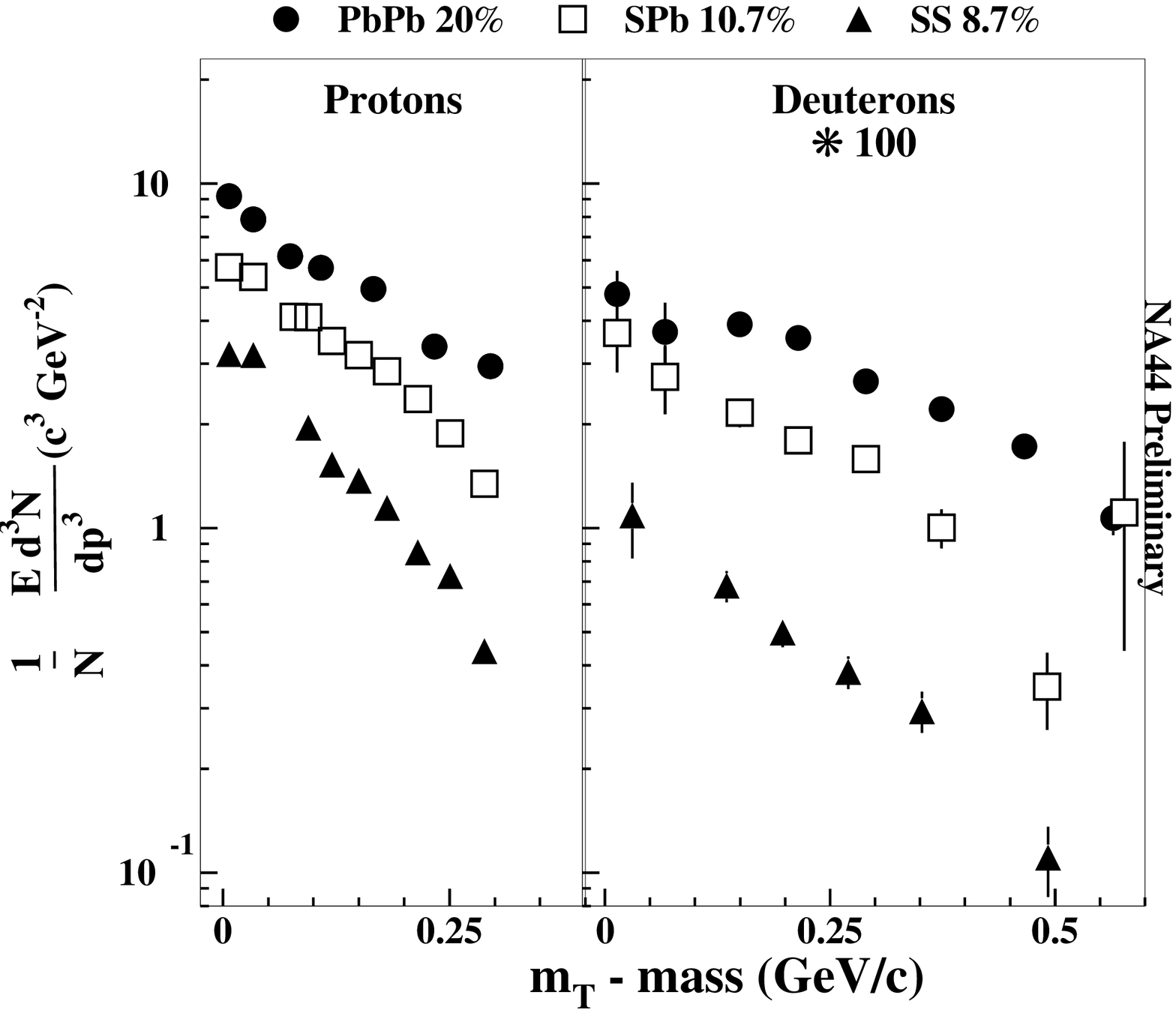}
      }
  \end{center}
  \vspace{-0.5cm}
\caption{Proton and deuteron $m_T$ spectra. The proton spectra have been 
corrected to remove protons from weak decays of $\Lambda^0$ and $\Sigma^+$.
 The systematic errors on
the normalization 
are 9,14 and 25\% for $SS$,$SPb$ and $PbPb$.}
\vspace{-0.5cm}
\label{mtspec}
  \vspace{-0.5cm}
 \begin{center}
    \mbox{
     \epsfxsize=10cm
     \epsffile{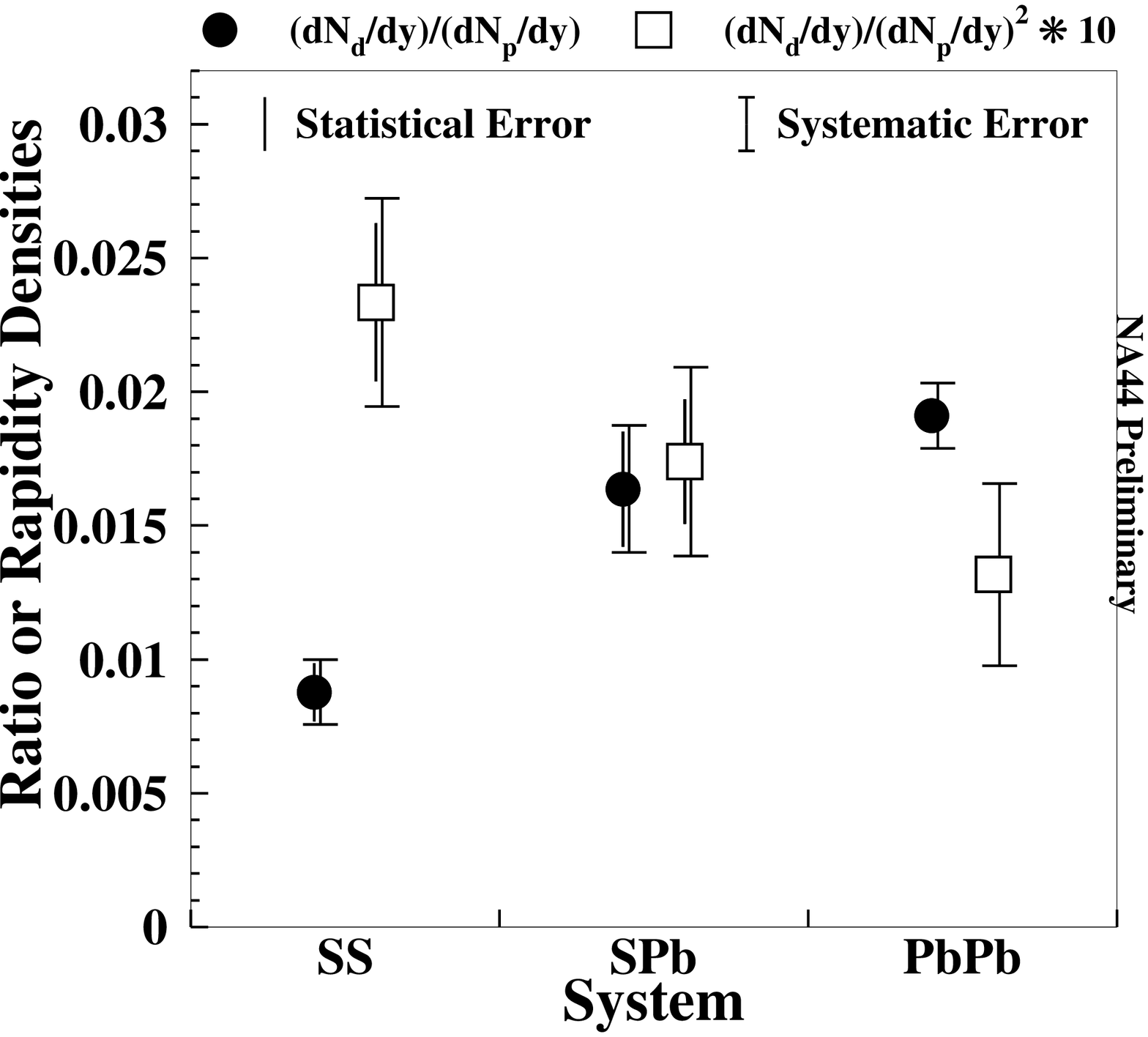}
      }
  \end{center}
  \vspace{-0.5cm}
\caption{$(dN_d/dy)/(dN_p/dy)$ and $(dN_d/dy)/(dN_p/dy)^2$ versus system}
\vspace{-0.5cm}
\label{summ}
\end{figure}

\begin{figure}[h]
 \vspace{-0.5cm}
 \begin{center}
    \mbox{
     \epsfxsize=10cm
     \epsffile{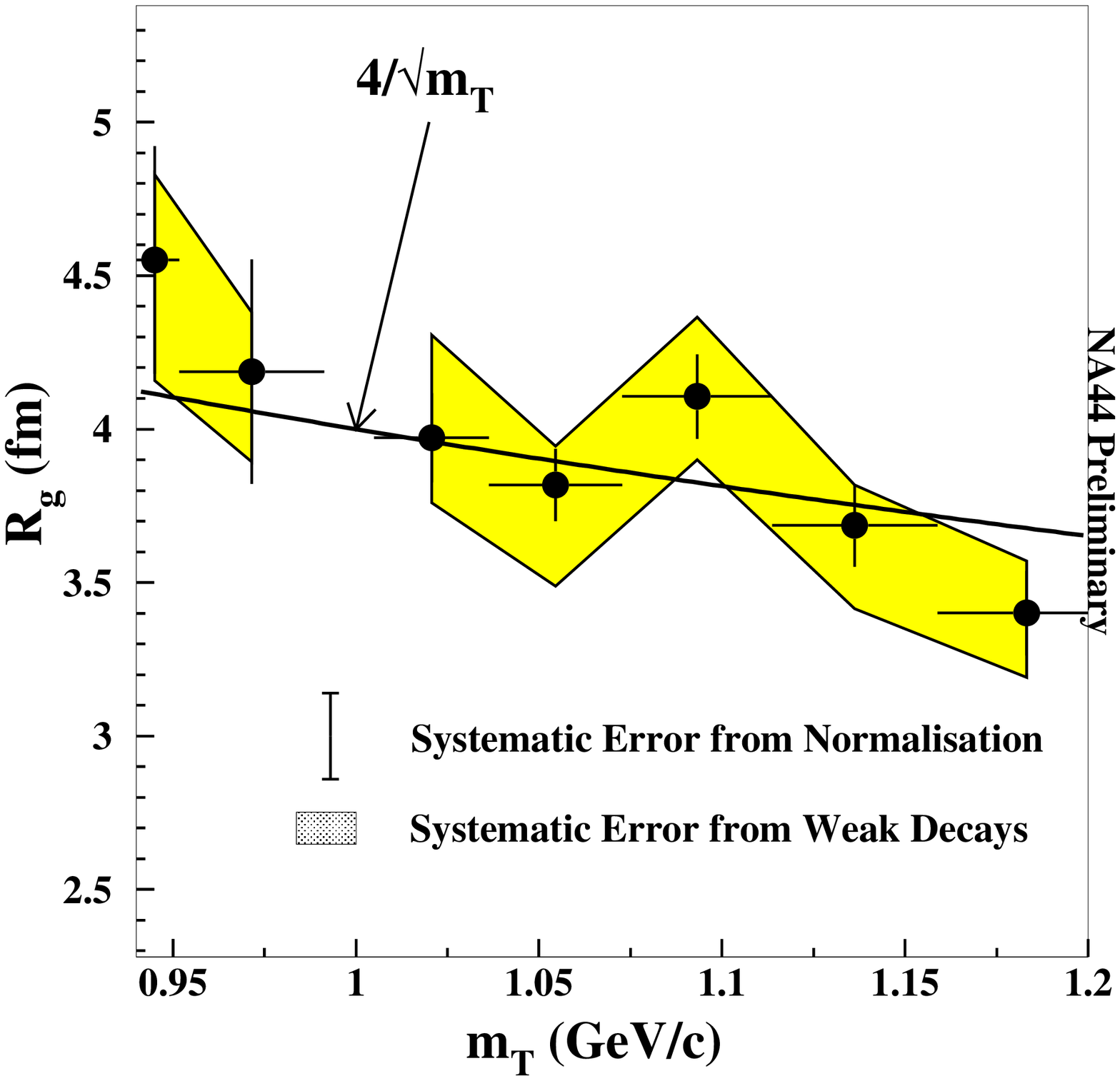}
      }
 \end{center}
  \vspace{-0.5cm}
\caption{$R_G$ versus the proton $m_T$ for $PbPb$ collisions.
Divide by $\protect{\sqrt{3}}$ to compare to 3D HBT radii from NA44.}
\vspace{-0.5cm}
\label{sqrtmt}
 \vspace{-0.5cm}
 \begin{center}
    \mbox{
     \epsfxsize=10cm
     \epsffile{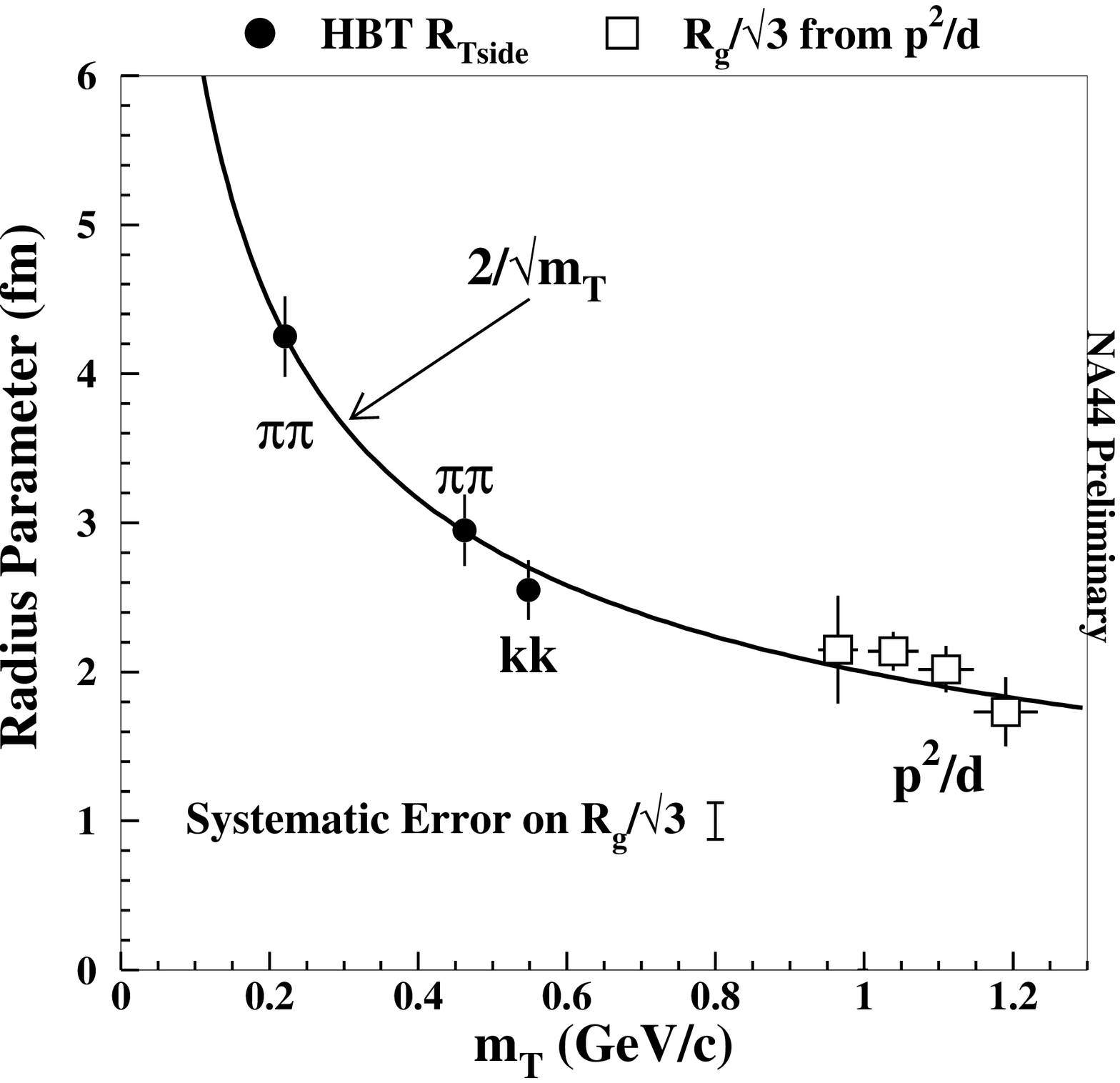}
      }
 \end{center}
  \vspace{-0.5cm}
\caption{HBT radii from [1] and $R_G/$$\protect{\sqrt{3}}$ 
versus $m_T$ for $SPb$.}
\vspace{-0.5cm}
\label{hbtdpp}
\end{figure}

Figure~
shows proton and deuteron $m_T$  spectra 
at y=1.9-2.3. 
Protons 
from the weak decays
of $\Lambda^0$ and $\Sigma^+$ are subtracted from the spectra 
using GEANT and RQMD, resulting in 
systematic errors $ \le 2\%$ on the slopes and  $\le 6\%$ on the dN/dy values.
The spectra flatten as the particle mass or the  system gets heavier as 
noted in  \cite{na4497a} for $\pi^\pm, K^\pm,p$ and $\bar{p}.~$
 For larger systems
more baryons are pushed into mid-rapidity. The energy they deposit 
produces particles that suffer secondary collisions
which may produce transverse flow and 
explain the trends in the slopes.
Figure \ref{summ}
 shows that 
$(dN_d/dy)/(dN_p/dy)$  increases from $SS$ to $PbPb$
while $(dN_d/dy)/(dN_p/dy)^2$ decreases.
In the coalescence model
deuterons form when $p-n$ pairs are close in phase space.
However, a 3rd particle must absorb the excess energy of the $p-n$ pair.
This is easier in $PbPb$ collisions.
In a spherically symmetric thermal model
,~\cite{note171},  with a gaussian  source
$\rho (r) = exp \frac{-r^2}{2R_G^2}$
the radius in the $p-n$ rest frame is 
\begin{equation}
R_G^3 = \frac{3}{4} (\pi )^{3/2} (\hbar c)^{3} \frac{m_d}{m_p^2} \cdot
\frac{(\frac{E_p \cdot d^3N_p}{dp^3})^2}
{\frac{E_d \cdot d^3N_d}{dp^3}}
\label{eq:RG_def}
\end{equation}
The  cross sections 
are evaluated at the same velocity for proton and deuterons and 
we assume $N_n = N_p$.
$R_G$ must be devided by $\sqrt{3}$ 
to compare it to NA44's 3d HBT radius parameters.
Figure~3 
shows that $R_G$ falls with $m_T$.
Collective
expansion creates position-momentum correlations
with correlation lengths 
$\propto 1/\sqrt{m_T}$, ~\cite{hydro1,hydro2}.
The data are consistent with this form.
Figure~\ref{hbtdpp} 
compares NA44 HBT radii, \cite{mtna44}, with $R_G/\sqrt{3}$ for $SPb$.
All $SPb$ radii agree with $2/\sqrt{m_T}$.

\section*{Conclusions}

Proton and deuteron $m_T$ spectra at $y$=1.9-2.3 are exponential 
with inverse 
slopes that increase with mass and system size.
The ratio $(dN_d/dy)/(dN_p/dy)$ increases 
From $SS$ to $PbPb$  but
$(dN_d/dy)/(dN_p/dy)^2$ decreases.
The radius parameter $R_G$ extracted from proton and deuteron 
spectra using Eqn. (1) 
increases from $SS$ to $PbPb$ and 
falls with $m_T$.
The increase of inverse slopes with mass
and the $1/\sqrt{m_T}$ dependence of $R_G$  
are consistent with radial flow. 

\end{document}